\documentstyle[aps,prl,twocolumn,psfig]{revtex}
\begin{document}


\newcommand{\tit}
{Signal of Quark Deconfinement \\ in the Timing
Structure of Pulsar Spin-Down}
\newcommand{\autha} {N. K. Glendenning}
\newcommand{\authb} {S. Pei}
\newcommand{\authc} {F. Weber}
\newcommand{\lbl}{\begin{flushright} LBL-39746\\[7ex] \end{flushright}}
\newcommand{\dateofdoc}{May, 12, 1997}
\newcommand{\doe}
{This work was supported by the
Director, Office of Energy Research,
Office of High Energy
and Nuclear Physics,
Division of Nuclear Physics,
of the U.S. Department of Energy under Contract
DE-AC03-76SF00098.}

\newcommand{\ect}{A part of this work was done at the ECT*,
Villa Tambosi, Trento, Italy.}

\begin{titlepage}
\lbl
\begin{center}
\begin{Large}
\renewcommand{\thefootnote}{\fnsymbol{footnote}}
\setcounter{footnote}{1}
\tit {\footnote{\doe}}\\[5ex]
\end{Large}

\renewcommand{\thefootnote}{\fnsymbol{footnote}}
\setcounter{footnote}{2}
\begin{large}
\autha {\footnote{\ect}},~~~ \authb~~~and~~~\authc \\[5ex]
\end{large}
\dateofdoc \\[5ex]
\end{center}


\begin{figure*}[tbh]
\vspace{-.5in}
\begin{center}
\leavevmode
\hspace{-.8in}
\psfig{figure=ps.no,width=4.in,height=4.in}
\end{center}
\end{figure*}

\begin{center}
\end{center}
\end{titlepage}

\clearpage
\draft
\title{Signal of Quark Deconfinement \\ in the Timing
Structure of Pulsar Spin-Down}
\author{Norman K. Glendenning}
\address{Nuclear Science Division and Institute for Nuclear and Particle
 Astrophysics,
 Lawrence Berkeley Laboratory,
 MS: 70A-3307, Berkeley, California 94720}
\author{S. Pei}
\address{Beijing Normal University,
Department of Physics,
Beijing 199875,
P. R. China}
\author{F. Weber}
\address{Ludwig-Maximilians University of Munich, Institute for Theoretical
Physics,
 Theresienstr. 37/III, 8000 Munich 2, Germany}

\date{\dateofdoc}
\maketitle

\begin{abstract}
The conversion of nuclear matter to quark matter in the core
of a rotating neutron star alters its moment of inertia. Hence the epoch
over which conversion takes place will be signaled in the spin-down
characteristics of pulsars. We find that an observable called the
braking index should be easily measurable during the transition
epoch and can have a value far removed (by orders of magnitude)
from the canonical
value of three expected for magnetic dipole radiation,
and may have either sign. The duration of the
transition epoch is governed by the slow loss of angular momentum
to radiation and is further prolonged by the reduction in the
moment of inertia caused by the phase change which can even
introduce an era of spin-up. We estimate that
about one in a hundred pulsars may be passing through this phase.
The phenomenon is analogous to ``bachbending'' observed in the moment
of inertia  of rotating nuclei  observed in the 1970's, which also
signaled a change in internal structure with changing spin.

\end{abstract}


The deconfined phase of hadronic matter called quark matter is believed to have
pervaded the early universe and may reside as a permanent component of neutron
stars in their dense high-pressure cores
\cite{baym76:a,chapline76:b,keister76:a,%
kislinger78:a,freedman78:a,joss78:a}.
However no means of detecting its presence has been found because the properties
of neutron stars  and those with a quark matter core are expected to be very 
similar. Alternately, instead of looking to the properties of the star itself,
we study the  spin-down behavior of a rotating star with the realization that
changes in internal structure as the star spins down  will be reflected in the
moment of inertia and hence in the deceleration.

Pulsars are born with an enormous store of angular momentum and
rotational energy which they radiate slowly over millions of
years by the weak processes of electromagnetic radiation and a
wind of electron-positron pairs
\cite{pacini67:a,goldreich69:a,ostriker69:a,ruderman87:a}. When
rotating rapidly, a pulsar is centrifugally flattened. The
interior density will rise with decreasing angular velocity and
may attain the critical density for a phase transition. First at
the center and then in an expanding region, matter will be
converted from the relatively incompressible nuclear matter phase
to the highly compressible quark matter phase. The weight of the
overlaying layers of nuclear matter will  compress the quark
matter core and the entire star will shrink. The mass
concentration will be further enhanced by the increasing
gravitational attraction of the core on the overlaying nuclear
matter. The moment of inertia thus decreases anomalously with
decreasing angular velocity as the new phase slowly engulfs a
growing fraction of the star.

\begin{figure}[h]
\vspace{-.7in}
\begin{center}
\leavevmode
\psfig{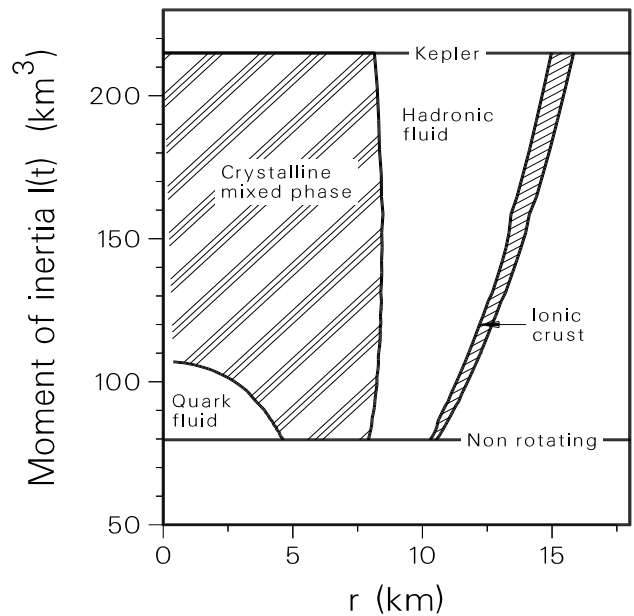}\\[.2in]
\parbox[t]{5.5 in} { \caption { \label{i_r_k300b180_d} Evolution  of
radial location between boundaries of various phases in a neutron star
of given baryon number as a function of moment of inertia, which is a
decreasing function of time.
Stellar mass is $1.42M_{\odot}$ for the nonrotating counterpart.
Formation of a pure quark matter core occurs below  $I\sim
107 ~{{\rm km^3}}~(\Omega
\approx 1400$ rad/s). Note the decrease of radius with time.
See Ref. \protect\cite{glen95:c} for a description of spatial geometry
in mixed phase.
}}
\end{center}
\end{figure}
The decrease of moment of inertia
caused by the phase transition
is superposed on   the
natural response of the stellar shape to a decreasing centrifugal
force occasioned by radiation loss. Therefore,
to conserve angular momentum not carried off by
radiation, the deceleration $\dot{\Omega}$ of the
angular velocity must respond by
decreasing  in absolute magnitude
and may actually change sign. The pulsar may
spin up for a time, just as an ice skater spins up upon contraction of the
arms before air resistance and friction
 reestablish spin-down.

\begin{figure}[tbh]
\vspace{-.4in}
\begin{center}
\leavevmode
\psfig{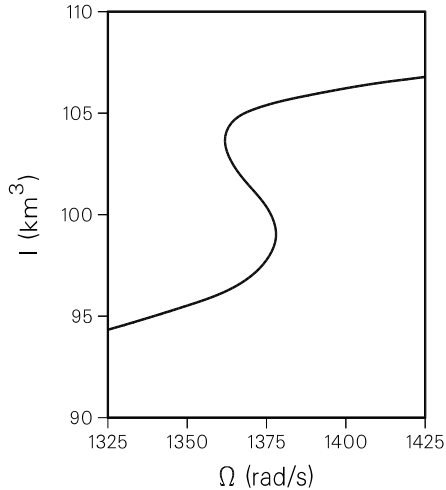}\\[1ex]
\parbox[t]{5.5in} { \caption { \label{oi} Moment of inertia
as a function of rotational angular velocity over
the epoch during which  an increasing central volume of the star
is passing into the deconfined quark matter phase. The temporal development
is from large to small $I$. Spin-up of the pulsar is clearly evident.
An analogous phenomenon known as ``backbending'' is observed in the rotational
properties of nuclei.
}}
\end{center}
\end{figure}

Such an anomalous
decrease of the moment of inertia as described  is analogous to the
phenomenon of ``backbending'' in the rotational bands of
nuclei  predicted by
Mottelson and Valatin and observed
years ago
\cite{mottelson60:a,johnson72:a,stephens72:a}. The moment of inertia
of a nucleus changes anomalously because
of  a change in phase from a nucleon spin-aligned phase at
high angular momentum
to a
pair-correlated superfluid phase at low.
The connection between the moment of inertia and the
internal structure of a neutron  star is shown in Fig.\ \ref{i_r_k300b180_d}
and between moment of inertia and frequency in Fig.\ \ref{oi}.

The property of asymptotic freedom of quarks makes a phase change
from confined to deconfined matter inevitable for sufficiently high
energy density of hadronic matter. We have therefore referred to the
transition as being from the confined to deconfined phase. However, it is
apparent in our discussion and results, that the
phenomenon   we  discuss
is not unique to this particular transition. It is simply
the most plausible in our view, but any phase transition
in dense hadronic matter is interesting.

Now we develop the appropriate
measure for detecting the occurrence of a
phase transition and its envelopment of an increasing
proportion of the mass and volume of the star with time.
The production of electromagnetic radiation  and a
wind of relativistic
pairs by the rotating magnetic field of the star exert a
torque on the star which is characteristic of magnetic dipole
radiation.
The  energy loss equation representing processes of multipolarity $n$
is of the
form
 \begin{eqnarray}
 \frac{dE}{dt} =
  \frac{d}{dt}\Bigl(\frac{1}{2} I \Omega^2 \Bigr) =
   - C \Omega^{n+1}
    \label{energyloss}
      \end{eqnarray}
      where $n=3$
        for  magnetic dipole radiation.
The rate of change of frequency is governed by
\begin{eqnarray}
\dot{\Omega}= -\frac{C}{I(\Omega)}
  \biggl[1  + \frac{I^{\prime}(\Omega) \,
\Omega}{2I(\Omega)}
 \biggr]^{-1} \Omega^n\,.
   \label{braking2}
\end{eqnarray}
For low frequency or
if changes
in $I$ are ignored
this reduces to the usual form quoted in the
literature. $\dot{\Omega}=-K \Omega^n$ where $K=C/I$,
(cf. Ref. \cite{manchester77:a,lyne90:b}).

The dimensionless measurable
quantity ${\Omega \ddot{\Omega} }/{\dot{\Omega}^2}$
is referred to  as the ``braking index''. It
would be  equal to the intrinsic index $n$
 of the energy-loss
 mechanism (\ref{energyloss})
 if the frequency were small or
if the moment of inertia were a constant.
However these conditions are not usually fulfilled and the measurable
quantity is not  constant. Rather it has the value
  \begin{eqnarray}
  n(\Omega)\equiv\frac{\Omega \ddot{\Omega} }{\dot{\Omega}^2}
  = n
   - \frac{ 3  I^\prime \Omega +I^{\prime \prime} \Omega^2 }
    {2I + I^\prime \Omega}
     \label{index}
      \end{eqnarray}
      where $I^\prime \equiv dI/d\Omega$ and $I^{\prime\prime}
      \equiv dI^2/d\Omega^2$.
The progression of the new phase through the central region
of the star will be signaled by an anomalous value of the braking
index, far removed from the canonical value of $n$.

Because the  pulsar
rotational energy  is coupled to weak processes,  $\dot{\Omega}$
is small and none of the quantities
in (\ref{index}) will change appreciably  over any observational
time span.
However the signal is carried in non-zero derivatives  $dI/d\Omega$
and $dI^2/d\Omega^2$;
 these are   large during the phase transition epoch
because of the progressive
conversion of
nuclear matter into
compressible quark matter.
As can be seen from (\ref{index}), large derivatives of the moment of inertia
will
 produce enormous deviations of the
 braking index from its canonical value
 while the region of the star occupied by
 quark matter is growing. Since the growth is paced by the slow  spin-down
 of the pulsar, we will find that the signal is ``on'' over an extended
epoch.

\begin{figure}[tbh]
\vspace{-.4in}
\begin{center}
\leavevmode
\psfig{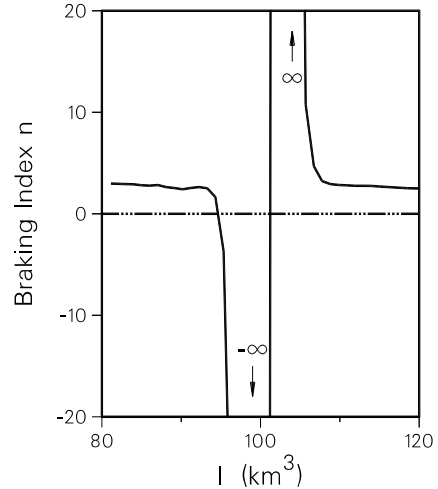}\\[1ex]
\parbox[t]{5.5in} { \caption { \label{ni} The
braking index as a function of moment of inertia signals by its large
departures from $n=3$ the progression of the deconfined phase in the
stellar interior. The braking index would be identically 3 if it were not for
the response of $I$ to rotation, and in any case tends to 3 for low
frequency (for magnetic dipole radiation).
}}
\end{center}
\end{figure}

The behavior of the moment of
inertia in the critical frequency interval
for our model star (which is described later) is shown in Fig.\ \ref{oi}.
As the pulsar evolves in time
(decreasing $I$) the  derivative   $dI/d\Omega$ passes
through two singularities, switching between
 $+\infty$ and  $-\infty$  at each turning point of
$\Omega$.
From
(\ref{braking2}) it is clear that the
 deceleration $\dot{\Omega}$ will pass through
 zero and change sign at both turning points becoming
an acceleration in the central part of the `S'; the pulsar spins up for a
time.
Moreover,
$-I^{\prime\prime}$
has similar singularities
as can be found from
\begin{eqnarray}
-I^{\prime\prime}=  \biggl( \frac{dI}{d\Omega}\biggr)^3
\frac{d^2\Omega}{dI^2}\,.
\end{eqnarray}
Consequently $n(\Omega)$
goes to $\pm \infty$ at the two turning points respectively
as shown in Fig.\ \ref{ni}.
We have plotted the braking index as a function of $I$
because $I$ decreases monotonically with the time, unlike $\Omega$.

(The spin-up referred  to above has nothing to do
with the minuscule spin-up known as  a pulsar glitch.
 The relative change in moment
 of inertia in a glitch episode is very small ($\Delta I/I \sim -\Delta \Omega
 /\Omega \sim 10^{-6}$ or smaller) and  approximates closely  a
 continuous response of the star
 to changing frequency on any time scale that is large
 compared to the glitch and recovery interval. Excursions of such a magnitude
 as quoted would fall within the thickness of the line
 in Fig.\ \ref{oi}.)

A  change in moment of inertia owing to a change in phase
such as we have described
is evidently a robust phenomenon. However, we cannot be sure that nature
will respond as strongly as our model does.
Backbending, as in Fig.\ \ref{oi}, is an extreme response
of the moment of inertia  to the progression of a phase
transition through the central region
of the star.
Instead, the transition of the moment of inertia
from that of a neutron star (at high $\Omega$) to a hybrid star may
be a
single-valued function of $\Omega$.
However the transition between these
two  moments
can still be marked by a large first derivative $I^\prime$ and a second
derivative
$I^{\prime\prime}$ that  is large and
changes sign in the transition interval. In this case, the braking index
will not swing between $\pm \infty$ but  nonetheless
can attain large positive
and negative values.

 Typically it is
difficult to measure  $\ddot{\Omega}$ (and hence the braking index)
because of timing noise.
 Only four braking indices are presently known. However,
for a star that is passing through the phase transition epoch,
the deceleration (\ref{braking2}) is reduced markedly (even
changing  to acceleration).
The second derivative must therefore be large in absolute magnitude
through all but the central portion of the epoch where spin-up
occurs.
Consequently,  the second derivative of frequency should be
easier to measure for pulsars passing through the epoch
than for typical pulsars.  Hence,  difficulty in measuring the
second derivative may be used as a de-selection criterion.

 We   estimate  the plausibility of observing
 in the pulsar population a signal of the
 kind that we find in our calculation.
 Assume that neutron stars are created in a narrow mass window
 (as present evidence suggests).
 The duration over which the observable index is
 anomalous is $\Delta T \approx -\Delta \Omega/\dot{\Omega}$ where
 $\Delta \Omega$ is the frequency interval of the anomaly.  The range in which
 $n$ is smaller than zero  or larger than six (Fig.\
 \ref{ni}) is $\Delta \Omega \approx 100$ rad/s.
 Take  a typical period derivative of
  $\dot{P}\sim 10^{-16}$ s/s to  find $\Delta T \sim 10^5$ years.
   The signal
     would endure for 1/100 of a typical   active pulsar lifetime.
     Similarly we
     estimate that spin-up would last for about 1/6'th of that time.
     Given that more than 700 pulsars
are presently known (200 of which have been discovered in the
last several years)
 about  7 of these  may be signaling the growth of a central
 region of new phase.

We describe briefly the calculation.
The equations describing the configurations of rotating
stellar structures are solved for a sequence having the same
baryon number but different rotational frequencies.
The usual expression for the moment of inertia
in General Relativity due to Hartle
is not adequate for our purpose
 (see Ref. \cite{hartle67:a}
 equations (49) and (62)). Not only are  effects of internal
constitutional changes  in the star absent but also absent are the effects
of
 the alteration of the metric of spacetime by rotation,
 the dragging of local inertial frames,
 and even  centrifugal flattening.
 Instead, to calculate the
 moment of inertia,  we must use an expression
 that incorporates the above  effects as derived by Glendenning and Weber
 \cite{glen92:b,glen93:a}.

The stellar  model
 is described as follows.
Neutron star matter has   a charge
neutral  composition of hadrons consisting of members of the
baryon octet together with leptons when in the purely confined phase.
The properties of such matter are calculated in a covariant mean field
theory as described in Refs.\
\cite{glen85:b,glen91:c,book,serot87:a,kapusta90:a,kapusta91:a}.
The values of the
parameters that define the coupling constants of the theory  are
certain fairly well constrained properties of nuclear matter
and hypernuclei as described in the references;
(binding energy of symmetric nuclear matter $B/A=-16.3$ MeV,
saturation density $\rho=0.153 {\rm ~fm^{-3}}$, compression modulus
$K=300$ MeV, symmetry energy coefficient $a_{{\rm sym}}=32.5$ MeV,
nucleon effective mass at saturation $m^{\star}_{{\rm sat}}=0.7m$ and
ratio of hyperon to nucleon couplings $x_\sigma=0.6,~x_\omega=0.653=x_\rho$
that yield, together with the foregoing parameters, the correct
$\Lambda$ binding in nuclear matter \cite{glen91:c}).
Quark matter
is treated in a version of the MIT bag model with the three light
flavor quarks
($m_u=m_d=0,~m_s=150$ MeV)
as described in Ref.\ \cite{glen91:a}.  A value of the
bag constant $B^{1/4}=180$ MeV is employed.
(See Refs. \cite{book,glen91:a,glen91:d}
for a correct treatment of first order phase transitions in multi-component
substances such as neutron star matter for which baryon and electric
charge are the
independent conserved charges.)
Very little is known about the high-density properties of matter and our
calculation does not imply
a prediction of the rotational {\sl frequency} or stellar
{\sl mass} at which a phase
transition will occur.
Rather it shows  what the signal could
be if conditions are attained  for the phase change.

So far as we know, there is little difference in the properties of
a neutron star that has no quark core and one that has already fully
developed one. But  a strong signal may be  associated with the gradual
conversion of matter from  one phase into the other
as the conversion
is paced by the slow loss of angular momentum
in the processes of electromagnetic radiation and electron-positron wind.
The   important observational features   by which
such an  epoch could be identified are:\\[1ex]
(1)  The braking index  has a value far from the canonical value,
possibly by orders of magnitude and can
be of either sign.\\[.5ex]
(2) The epoch over which the braking index is anomalous is long because
pulsars spin down slowly under ordinary circumstances but even
more slowly  when $|I^\prime|$ is large (see (\ref{braking2})).\\[.5ex]
(3)
The pulsar may be observed to be spinning up. (To avoid confusion
with spin-up due to accretion, only isolated pulsars are relevant.)\\[.5ex]
(4)
Except for the central part of the spin-up era, the derivative
$\ddot{\Omega}$ is large and therefore easy to measure and so also is
the braking index.\\[.5ex]
(5)
Difficulty in measuring $\ddot{\Omega}$ can be used to deselect
phase transition candidates.\\[.5ex]
(6)
An estimated 1/100 pulsars may be passing through the transition epoch.\\[.5ex]

Pulsar observations are still in their infancy. It takes a considerable
time-span of data to measure the braking index. And many of the presently
known pulsars have only recently been discovered. We conclude from
our work that it is
plausible that the phase transition signal can be observed. It would be a
momentous discovery to find that a phase of matter that existed in the
early universe inhabits the cores of some neutron stars.\\[1ex]
       \doe \\ \indent \ect~One of us (NKG) is grateful to the hospitality of
       the center.


\end{document}